\documentclass[preprint,aps,showpacs,nofootinbib,preprintnumbers,amsmath,amssymb]{revtex4-1}
\usepackage{amssymb}
\usepackage{epsfig}
\usepackage{graphicx}
\usepackage{subfigure}
\usepackage{dcolumn}
\usepackage{bm}
\usepackage[usenames ,dvipsnames]{xcolor}
\usepackage{slashed}

\begin{document}

\title{Stochastic Gravitational Waves from Inflaton Decays}

\author{Da~Huang}\email{dahuang@fuw.edu.pl}
 \affiliation{Institute of Theoretical Physics, Faculty of Physics, University of Warsaw, Pasteura 5, 02-093 Warsaw, Poland}

\author{Lu Yin}\email{yinlu@gapp.nthu.edu.tw}
\affiliation{Department of Physics, National Tsing Hua University, Hsinchu, Taiwan 300}

\date{\today}
\begin{abstract}
Due to the universality of gravitational interactions, it is generally expected that a stochastic gravitational wave (GW) background could form during the reheating period when the inflaton perturbatively decays with the emission of gravitons. Previously, only models in which the inflaton dominantly decays into a pair of light scalar and/or fermion particles were considered in the literature. In the present paper, we focus on the cases with a vector particle pair in the final decay product. The differential decay rates for the three-body gravitational inflaton decays are presented for two typical couplings between the inflaton and vector fields, from which we predict their respective GW frequency spectra. It turns out that, similar to the scalar and fermion cases, the obtained GW spectra is too high in frequency to be observed by the current and near-future GW detection experiments and calls for a new design of high-frequency GW detectors.
\end{abstract}

\maketitle

\section{Introduction}
After the discovery of gravitational waves (GWs) by the advanced LIGO (aLIGO) Collaboration~\cite{Abbott:2016blz}, we are entering an exciting era to exploit the GWs to probe the early Universe and new physics beyond the Standard Model (SM). In the literature, except for the primordial vacuum tensor fluctuations during the inflation, most traditional sources produce the GW in a classical manner~\cite{Binetruy:2012ze, Maggiore}, even though there have been several proposals for GWs of particle origin during the inflation~\cite{Senatore:2011sp}.

Recently, it was proposed in Ref.~\cite{Nakayama:2018ptw} that one possible source to the stochastic GW background might come from heavy particle decays in the early Universe. This mechanism is particularly interesting if we consider the decays of the inflaton during the reheating epoch. After the end of the inflation, the inflaton field would roll down the potetial quickly and coherently oscillate around its bottom. This coherent inflaton oscillation behaves like a non-relativistic matter due to its vanishing pressure and leads to a period of matter domination. In the case that the nonlinear preheating process is inefficient or does not occur at all~\cite{Chung:1998rq}, the reheating would proceed via the perturbative inflaton decays into some lighter particles, whose further decays or scatterings with other SM particles could lead to the thermalzation of the SM sector. In the light of the universality of the gravitational interaction, it is unavoidable that gravitons can be emitted by the inflaton decays but with a rate suppressed by a factor of $(M/M_{\rm Pl})^2$ in which $M$ denotes the mass of inflaton and $M_{\rm Pl}$ the reduced Planck mass. As a result, it was found in Ref.~\cite{Nakayama:2018ptw} that when the inflaton mass was of ${\cal O}(M_{\rm Pl})$, a fraction of ${\cal O}(10^{-2}) $ of the whole inflaton field energy could be carried away by gravitons, which would ultimately form a stochastic GW background after redshifting with the cosmic expansion. Unfortunately, the obtained GW spectrum was typically too high in frequency to be observed by the current and near-future GW detection experiments.

Note that the authors in Ref.~\cite{Nakayama:2018ptw} only considered the models in which the inflaton decay was dominated by processes with a pair of scalar or fermion particles in the final states. In order to fully explore this mechanism, we should consider different types of particles and their various interactions. In the present paper, we focus on GW productions in models in which the inflaton mainly decays into a pair of vector particles during reheating. We are interested to see if there is some new features in the obtained GW spectra and if we could decrease the typical GW frequencies to be within the sensitivity regions of the present and future GW experiments.

The paper is organized as follows. In Sec.~\ref{SecModel}, we present our models in which the inflaton dominantly decays into a pair of vector particles. We consider two types of inflaton-vector couplings, from which we derive their respective differential graviton energy spectra of the three-body inflaton gravitational decays. Then we calculate the typical spectra of the stochastic GW backgrounds for these two cases in Sec.~\ref{SecGW}. In the calculations, we pay attention to the neutrino decoupling effect in the final formula of the GW spectrum, which was ignored earlier in Ref.~\cite{Nakayama:2018ptw}. Finally, in Sec.~\ref{SecConc}, we present our conclusion and some further discussion. 

\section{Partial Decay Rates from Inflaton Gravitational Decays}\label{SecModel}
In this section, we present our models and the graviton spectra from the gravitational inflaton decays. The general action describing the interactions of the inflaton $\sigma$, the graviton and a massive vector field $A$ can be written in the Einstein frame as follows
\begin{eqnarray}
S = \int d^4 x \sqrt{|g|} \left[\frac{M_{\rm Pl}^2}{2}R + \frac{1}{2}g^{\mu\nu} \partial_\mu \sigma \partial_\nu \sigma - V(\sigma) -\frac{1}{4} g^{\mu\rho}g^{\nu\sigma} F_{\mu\nu}F_{\rho\sigma} - \frac{m_A^2}{2} g^{\mu\nu} A_\mu A_\nu -\delta {\cal L} \right]\,,\nonumber\\
\end{eqnarray}
where $M_{\rm Pl} \equiv 1/\sqrt{8\pi G} = 2.44\times 10^{18}$~GeV is the reduced Planck mass, $F_{\mu\nu}$ denotes the field strength for $A$, and $V(\sigma)$ represents the inflaton potential. Since we are interested in the inflaton decay processes when reheating, only the potential near its minimum is involved in the calculation and can be approximated as $V(\sigma) =  M^2 \sigma^2/2$ with $M$ the inflaton mass. $\delta {\cal L}$ describes the interaction between the inflaton $\sigma$ and the vector field $A$. In the following, we shall consider two possible interactions:
\begin{eqnarray}\label{IntH}
\delta {\cal L}^{H} = \frac{\mu}{2} g^{\mu\nu} \sigma A_\mu A_\nu \,,
\end{eqnarray}
and
\begin{eqnarray}\label{IntA}
\delta {\cal L}^{A} = \frac{1}{f} \sigma \tilde{F}^{\mu\nu}  F_{\mu\nu} \,,
\end{eqnarray}
where $\tilde{F}^{\mu\nu} = (1/2) \epsilon^{\mu\nu\rho\sigma} F_{\rho\sigma}$ denotes the dual of the field strength. Note that the former interaction can naturally arise in a theory where the gauge symmetry for the vector field $A$ is spontaneously broken by the Higgs mechanism, while in the latter one the inflaton behaves like an axion when coupled to $A$. Thus, for simplicity, we will call these two interactions as Higgs-like and axion-like respectively, which also explain the superscripts in Eqs.~(\ref{IntH}) and (\ref{IntA}).

With the two interactions above, we can easily obtain their respective two-body inflaton decay rates as follows
\begin{eqnarray}
\Gamma_0^H (\sigma \to A A) &=& \frac{M}{64\pi} \left(\frac{\mu}{M}\right)^2 \frac{1}{y^4} \left(1- 4 y^2\right)^{3/2}\,,\label{Rate0H}\\
\Gamma_0^A (\sigma \to AA) &=& \frac{M}{4\pi}\left(\frac{M}{f}\right)^2 \left(1-4y^2\right)^{3/2}\label{Rate0A}\,,
\end{eqnarray}
where we have defined $y\equiv m_A/M$. It is seen that the inflaton decay rate for the Higgs-like coupling $\Gamma^H_0$ is divergent in the limit $y\to 0$, indicating that there is not massless limit in this case. In contrast, the axion-like coupling leads to a finite decay rate when $y \to 0$.

\begin{figure}[th]
\includegraphics[width = 0.95 \linewidth]{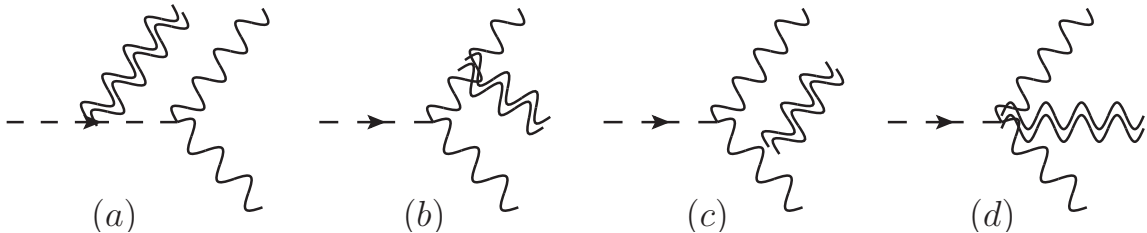}
\caption{Feynman Diagrams relevant to the three-body decays of the inflaton (dashed line) to a pair of vector particles (singly wiggle line) and a graviton (doubly wiggle line).}\label{FigFeyn}
\end{figure}

In order to compute three-body inflaton decays with graviton emissions, we need to decompose the metric tensor field into the flat background and the quantum fluctuation as $g_{\mu\nu} = \eta_{\mu\nu}+\kappa h_{\mu\nu}$ with $\kappa \equiv \sqrt{16\pi G} = \sqrt{2}/M_{\rm Pl}$, and expand the Lagrangian to the leading order of the perturbation $h_{\mu\nu}$. As a result, the graviton interactions with other particles are given by:
\begin{eqnarray}
\delta {\cal L} \supset \frac{\kappa}{2} h_{\mu\nu} T^{\mu\nu}\,,
\end{eqnarray}
where $T^{\mu\nu}$ is the energy-momentum tensor of matter fields, $\sigma$ and $A$. For the resultant Feynman rules, we apply those listed in Ref.~\cite{Han:1998sg}, from which we can draw the relevant Feynman diagrams shown in Fig.~\ref{FigFeyn} for the inflaton gravitational decays.
Note that the Feynman diagram (d) in Fig.~\ref{FigFeyn} is absent for the axion-like coupling since this interaction does not have any dependence on the metric tensor. By the tedious but straightforward computations, we can yield the expressions for the graviton energy spectrum in this process:
\begin{eqnarray}\label{DecayRate1H}
\frac{d\Gamma^H_1}{Mdx} &=& \frac{1}{64\pi^3} \left(\frac{\mu}{M_{\rm Pl}}\right)^2 \frac{1}{32 x y^4} \Bigg\{ \left[ 1 -4 x + 4x^2 - 2y^2 + 12 x y^2 - 48 x^2 y^2 \right. \nonumber\\
&& \left.  + 64 x^3 y^2 + 4 y^4 - 32 x y^4 + 48 x^2 y^4 + 24 y^6 - 48 x y^6 \right]\alpha \nonumber\\
&& - 4 y^2 \left[ 1 - 2x -4 x^2 + 8 x^3 - 5 y^2 + 8 x y^2 \right. \nonumber\\
&& \left. + 16 y^4 - 24 x y^4 -12 y^6 \right] \ln \left(\frac{1+\alpha}{1-\alpha}\right) \Bigg\}\,,
\end{eqnarray}
for the Higgs-like coupling $\delta{\cal L}^H$, and
\begin{eqnarray}\label{DecayRate1A}
\frac{d\Gamma^A_1}{Mdx} &=& \frac{1}{64\pi^3} \left(\frac{M}{f}\right)^2 \left(\frac{M}{M_{\rm Pl}}\right)^2 \frac{1}{x} \Bigg\{ \left[1 - 4x + 12 x^2 - 16 x^3 + 8 x^4 \right.\nonumber\\
&& \left. -2 y^2 +12 xy^2 - 16 x^2 y^2 - 8 y^4 + 16 xy^4 \right] \alpha  \nonumber\\
&& -4 y^2 \left[ 1 - 2 x + 6x^2 - 4 x^3 - 5 y^2 + 8 x y^2 - 8 x^2 y^2 + 4 y^4  \right] \ln \left(\frac{1+\alpha}{1-\alpha}\right) \Bigg\}\,,
\end{eqnarray}
for the axion-like coupling $\delta {\cal L}^A$,
where we have defined the following variables for simplicity:
\begin{eqnarray}
x = E/M\,,\quad \quad \alpha = \sqrt{1-\frac{4y^2}{1-2x}}\,.
\end{eqnarray}
Here we always assume that the decay products $A$ are much lighter than the inflaton {\it i.e.}, $y \ll 1$. In this limit, the dependence of $d\Gamma^A_1/(M dx)$ for the axion-like coupling on $y$ is seen to be very weak, while the decay rate of the Higgs-like coupling is divergent as $1/y^4$, which is similar to its associated two-body process in Eq.~(\ref{Rate0H}). However, as shown later, such a singular behavior would be cancelled in the final prediction of gravitational wave observables. Thus, for illustration, we fix $y=0.1$ in the following.

Furthermore, for a fixed $y$, it is easy to see that the differential decay rate $d\Gamma^{H,A}_1/(Mdx)$ is divergent as $1/x$ in the low graviton energy limit $x\to 0$ for both types of inflaton-vector interactions. In order to yield a sensible finite decay rate $\Gamma^{H,A}_1$, we need to regularize the integrations in the small $x$ region. This is done practically by introducing an infrared (IR) cutoff scale $\Lambda$ for the radiated graviton energy $E$, which is transformed into the lowest integration limit of $x_{L} = \Lambda/M$.

\begin{figure}[th]
\includegraphics[scale = 0.4]{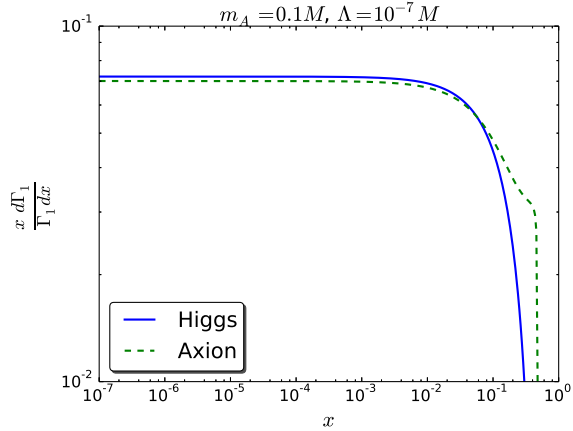}
\caption{Normalized graviton spectra $xd\Gamma^{H,A}_1/(\Gamma^{H,A}_1 dx)$ in the three-body inflaton gravitational decays when $\Lambda = 10^{-7} M$. The blue solid and green dashed curves denote the spectra for the Higgs-like and axion-like inflaton-vector interactions, respectively. }\label{FigVecSpec}
\end{figure}

In Fig.~\ref{FigVecSpec}, we plot the normalized graviton spectrum $x\,d\Gamma^{H,A}_1/(\Gamma_1 dx)$ for both the Higgs-like (blue solid curve) and axion-like (green dashed curve) inflaton-vector interactions where the IR energy cutoff is taken to be $\Lambda = 10^{-7} M$.
According to its definition, this graviton spectrum is well-defined in the small $y$ region for the Higgs-like coupling since the singular $1/y^4$ behavior in Eq.~(\ref{DecayRate1H}) is totally cancelled out. Furthermore, it is seen that both spectra approach constant values when $x\to 0$. This observation reflects the fact that the partial gravitational decay rates have $1/x$ behavior in this parameter region, indicating that the radiated gravitons are mostly concentrated in the low energy.

\begin{figure}[ht]
\includegraphics[width=0.47\linewidth]{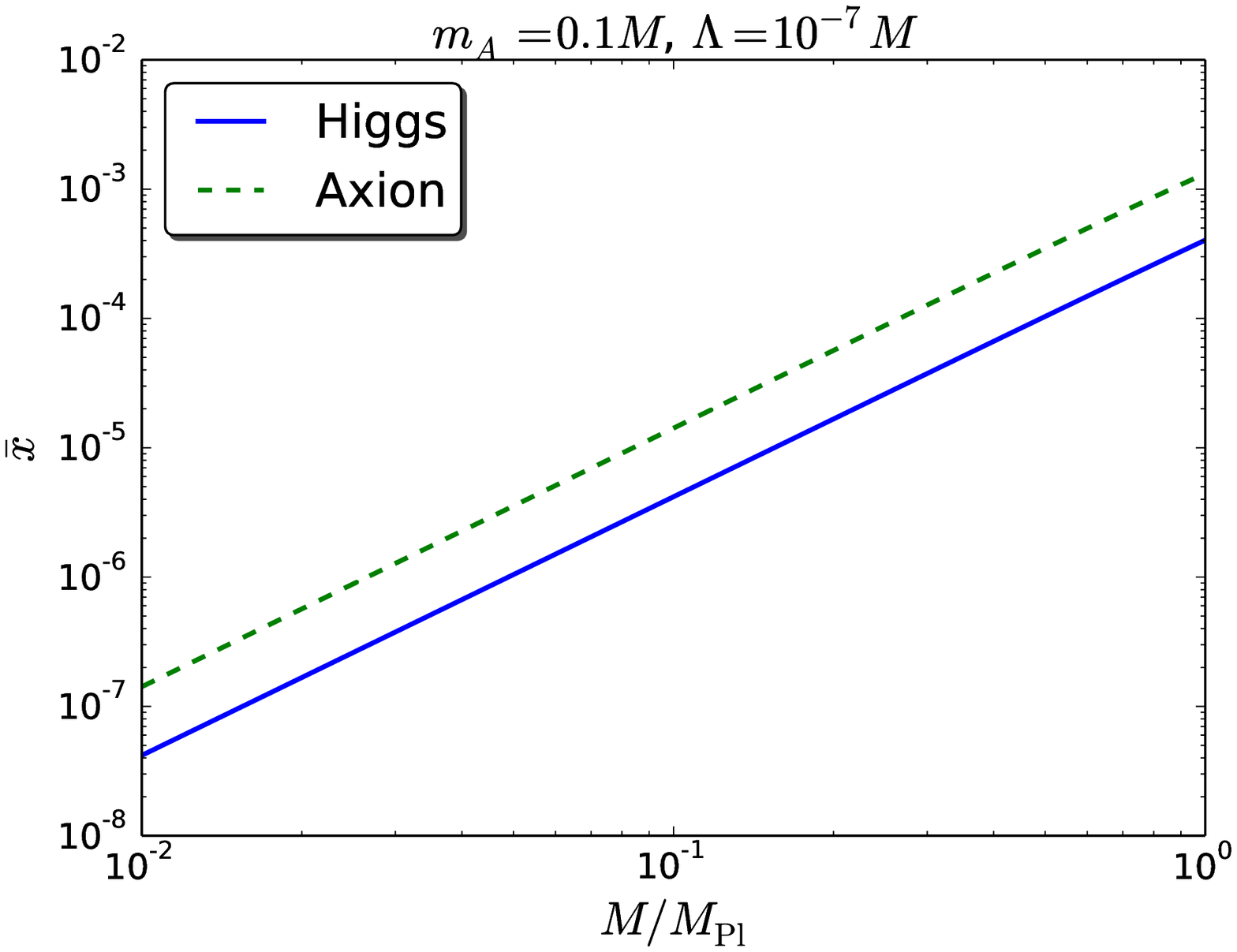}
\includegraphics[width = 0.47 \linewidth]{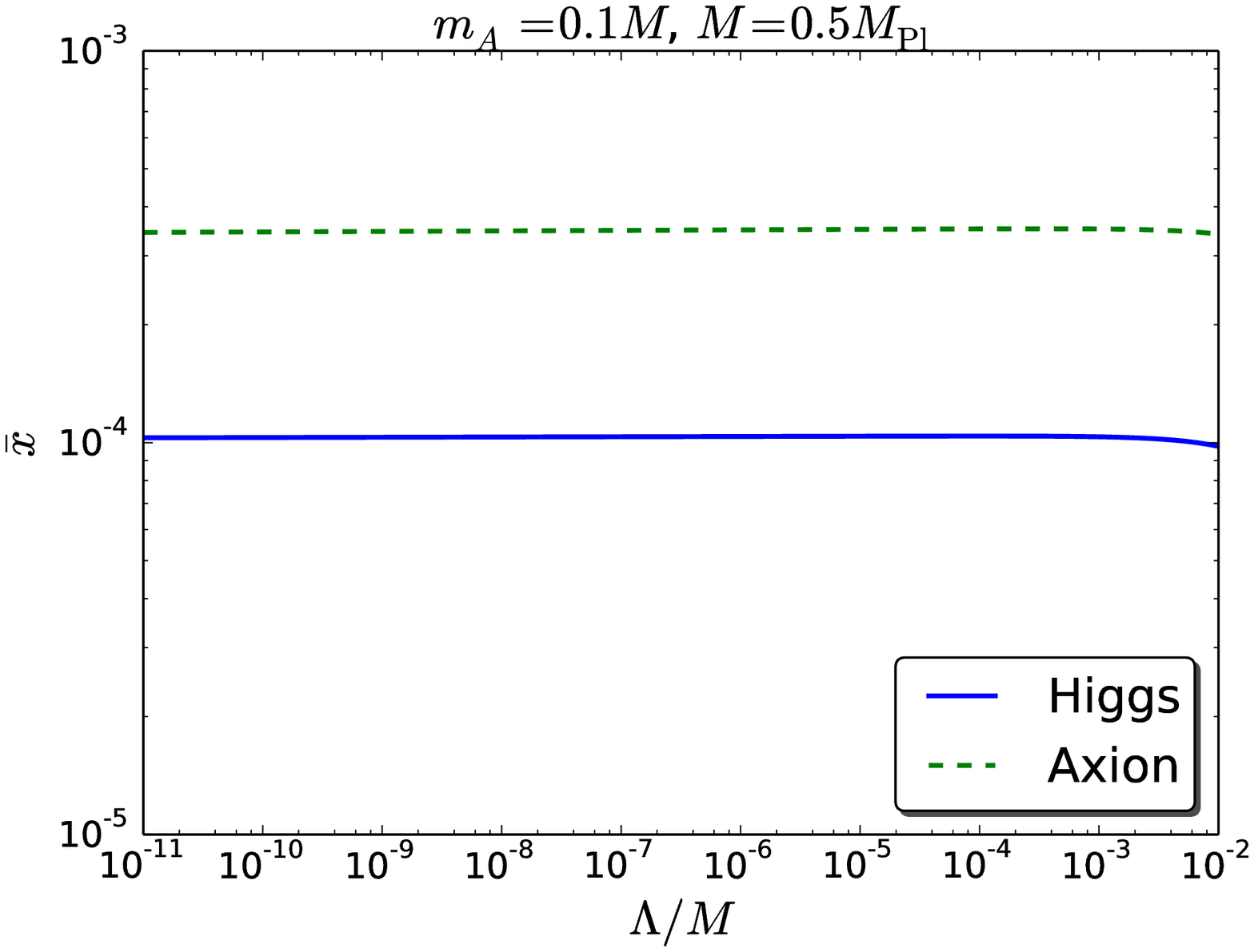}
\caption{Fraction of energy carried away by gravitons during inflaton decays, $\bar{x}$, as a function of the inflaton mass $M/M_{\rm Pl}$ (left panel) and the IR cutoff scale $\Lambda/M$ (right panel).}\label{FigVecXbar}
\end{figure}

Another important quantity characterizing the GW production during reheating is the energy fraction carried away by gravitons in inflaton decays, which is defined as follows:
\begin{eqnarray}
\bar{x} \equiv \frac{\bar{E}}{M} = \frac{\Gamma_1}{\Gamma}\int^{x_M}_{x_L} \frac{x d\Gamma_1}{\Gamma_1 dx} dx \,,
\end{eqnarray}
where $\Gamma =\Gamma_0+\Gamma_1$ denotes the total decay rate of the inflaton, while $x_M$ ($x_L$) denotes the largest (lowest) energy fraction that can be taken away by a graviton in a single three-body decay process. $x_M = (1-4y^2)/2$ is determined by the three-body decay kinematics, and $x_L = \Lambda/M$ is given by the IR cutoff scale. Similar to the normalized energy spectrum, $\bar{x}$ is finite in the limit $y \to 0$ for the Higgs-like coupling. In Fig.~\ref{FigVecXbar}, we show $\bar{x}$ as a function of the inflaton mass $M$ (Left Panel) and the IR cutoff $\Lambda$ (Right Panel) for both interactions.
It is seen from the left panel of Fig.~\ref{FigVecXbar} that $\bar{x}$ increases as $M^2$ in the whole inflaton mass range of interest. Moreover, when $M$ approaches the Planck scale, both graviton energy fractions can be as large as ${\cal O}(10^{-3})$. On the other hand, the right panel of Fig.~\ref{FigVecXbar} shows that the prediction of $\bar{x}$ for either coupling is insensitive to the modification of the IR cutoff scale $\Lambda$ when $\Lambda < 10^{-2} M$.

\section{Stochastic Gravitational Wave Spectrum}\label{SecGW}
After produced via inflaton decays in the reheating process, gravitons would propagate and spread in the whole Universe without any further interactions with other particles. As a result, it is expected that they would form a homogeneous and isotropic stochastic GW background at present after the attenuation of their energies and amplitudes due to the cosmic expansion. The basic observable for this GW background is the following partial energy density fraction:
\begin{eqnarray}
\Omega_{\rm GW}(f) \equiv \frac{1}{\rho_c^0} \frac{d \rho_{GW}}{d \ln f}\,,
\end{eqnarray}
where $f$ denotes the frequency of the GW signal which is related to its energy $E_0$ as $f = E_0/(2\pi)$ and $\rho_c^0 = 3 M_{\rm Pl}^2 H_0^2$ is the present critical energy density.

\subsection{Analytic Derivation of the GW Spectrum}
In order to proceed, we may rewrite $\Omega(f)$ in the following form
\begin{eqnarray}\label{omega2}
\Omega(f) = \Omega_\gamma \left[\frac{d (\rho_{\rm GW}/\rho_\gamma^0)}{d\ln E_0}\right]\,,
\end{eqnarray}
where $\Omega_\gamma = \rho^0_\gamma/\rho_c = 5.38\times 10^{-5}$ is the energy density parameter of photons today. 

Now we show that the factor in the square bracket of Eq.~(\ref{omega2}) can be related to the partial rates $d\Gamma_1/(M dx)$ of the three-body inflaton gravitational decays. Firstly, we assume that the reheating is completed instantaneously after the inflaton decay, so that the Hubble parameter then should be equal to the total inflaton decay rate $H=\Gamma$. By applying the energy conservation, the reheating temperature can be estimated as
\begin{eqnarray}\label{Tre}
T_R = \left[\frac{90}{\pi^2 g_\rho(T_R)}\right]^{1/4}\sqrt{M_{\rm Pl}\Gamma} = 0.54 \sqrt{M_{\rm Pl}\Gamma} \,,
\end{eqnarray}
which we have used the Friedmann equation. In the second equality, we take the relativistic degrees of freedom (dofs) in the plasma to be $g_\rho (T_R) = 106.75$, which assumes that the reheating temperature is so high that all SM particles are thermalized. Caused by the expansion of the Universe, both the frequency and amplitude of the GW signal of reheating are redshifted. Concretely, the GW energy simply evolves as
\begin{eqnarray}
E_0 = E_R \left(\frac{a_R}{a_0}\right)\,,
\end{eqnarray}
where the subscripts ``$R$'' (``0'') represent the values of the corresponding quantities at the reheating (present) time. According to the entropy conservation in the unit comoving volume, we can obtain the expansion factor $a_R/a_0$. Before the neutrino decoupling, all of the SM particles are in thermal equilibrium, so that we have the relation $g^R_s T_R^3 a_R^3= g_s^{\nu\,a} T_\nu^3 a_\nu^3$, in which $g_s^{R} = 106.75$ ($g_s^{\nu\,a} = 43/4$), $T_{R(\nu)}$ and $a_{R(\nu)}$ denote the relativistic dofs, temperature, and scale factor at reheating (just before neutrino decoupling), respectively. Nevertheless, after the neutrino decoupling, only electrons and positrons are equilibrated with photons, so that the dofs drops down to $g_s^{\nu\,b} = 11/2$. The subsequent entropy conservation gives $g_s^{\nu\,b} T_\nu^3 a_\nu^3 = g_s^0 T_0^3 a_0^3$. By combining the previous two evolutions, we can obtain
\begin{eqnarray}\label{aExp}
\frac{a_R}{a_0} = \frac{T_0}{T_R} \left(\frac{g_s^0}{g_s^R} \right)^{1/3} \left( \frac{g_s^{\nu a}}{g_s^{\nu b}}\right)^{1/3}\,,
\end{eqnarray}
which is different from the formula in Ref.~\cite{Nakayama:2018ptw} in the last factor which accounts for the neutrino decoupling effect.

On the other hand, since gravitons are massless, their energy density changes according to $\rho_{\rm GW}^0 = (a_R/a_0)^4 \rho_{\rm GW}^R$, while today's photon energy density can be written as follows
\begin{eqnarray}
\rho_\gamma^0 = \frac{\pi^2}{30} g_\rho^0 T_0^4 = \frac{g_\rho^0}{g_\rho^R} \left(\frac{T_0}{T_R}\right)^4 \rho_\gamma^R\,,
\end{eqnarray}
where the relativistic dofs in energy density are $g_\rho^R = g_s^R = 106.75$ ($g_\rho^0 = g_s^0 = 2$) at reheating (present). Therefore, taking the ratio of these two equations yields
\begin{eqnarray}
\frac{\rho_{\rm GW}^0}{\rho_\gamma^0} = \frac{\rho_{\rm GW}^R}{\rho_\gamma^R}\left(\frac{g_\rho ^R}{g_\rho^0}\right) \left(\frac{a_R}{a_0}\right)^4 \left(\frac{T_R}{T_0}\right)^4 = \frac{\rho_{\rm GW}^R}{\rho_\gamma^R} \left(\frac{g_s^0}{g_s^R} \right)^{1/3} \left( \frac{g_s^{\nu a}}{g_s^{\nu b}}\right)^{4/3}\,,
\end{eqnarray}
where we have used Eq.~(\ref{aExp}) to reduce the expression. Note that $\rho_{\rm GW}^R/\rho_\gamma^R$ is the ratio of the GW and SM plasma energy densities at the reheating temperature, whose energy spectrum can be approximated as
\begin{eqnarray}
\frac{d (\rho_{\rm GW}^R/\rho_\gamma^R)}{d\ln E_R} = \frac{1}{1-\bar{x}} \frac{x^2 d\Gamma_1}{\Gamma dx}\,,
\end{eqnarray}
in which $x = E_R/M$. By putting all the factors together, we can obtain
\begin{eqnarray}\label{GWSpec}
\Omega_{\rm GW}(f) = \Omega_\gamma \left(\frac{g_s^0}{g_s^R} \right)^{1/3} \left( \frac{g_s^{\nu a}}{g_s^{\nu b}}\right)^{4/3} \frac{1}{1-\bar{x}} \frac{x^2 d\Gamma_1}{\Gamma dx} \,,
\end{eqnarray}
where $x$ can be written as a function of the present GW frequency as follows
\begin{eqnarray}\label{xRed}
x = \frac{E_R}{M} = \left(\frac{a_0}{a_R}\right) \frac{E_0}{M} =  \left(\frac{T_R}{T_0}\right) \left(\frac{g_s^R}{g_s^0} \right)^{1/3} \left( \frac{g_s^{\nu b}}{g_s^{\nu a}}\right)^{1/3} \frac{2\pi f}{M}\,.
\end{eqnarray}
It is easy to see that the final GW spectrum in Eq.~(\ref{GWSpec}) should be well-defined in the limit of $y \to 0$ for the Higgs-like inflaton interaction since it only depends on $y$ through $\bar{x}$ and $d\Gamma_1/(\Gamma dx)$, both of which are finite. Moreover, due to the fact that $\bar{x}$ and $d\Gamma_1/(\Gamma dx)$ depend on $y$ at most logarithmically, the GW spectrum is insensitive to the precise value of $y$ as long as $y\ll 1$, so that the predictions based on $y=0.1$ is quite generic.

\subsection{Numerical Calculations of GW Signals}
As an application of Eq.~(\ref{GWSpec}), we can predict the expected GW spectrum produced by three-body gravitational decays of the inflaton during reheating. In the following, we take the instantaneous reheating approximation, in which after the decay of inflaton to the vector pair with a rate $\Gamma$, light SM particles can be generated and thermalized soon via the further decays or/and annihilations of $A$. Thus, the Hubble parameter at reheating is $H=\Gamma$, and the reheating temperature is given as in Eq.~(\ref{Tre}). As shown in Ref.~\cite{Nakayama:2018ptw}, a more detailed calculation by taking into account the finite decay time effect does not affect the final results much. Another important issue in our calculation of GW spectrum is the determination of the IR cutoff scale $\Lambda$ in order to obtain a well-defined decay rate. Here we take the cutoff scale to be the Hubble parameter $\Lambda = H$ during reheating, rather than the energy scale derived from the average inflaton number density used in Ref.~\cite{Nakayama:2018ptw}. Under these assumptions, the whole inflaton decay can be characterised by only two free parameters: the inflaton mass $M$ and its total decay rate $\Gamma$.

\begin{figure}[ht]
\includegraphics[width=0.49\linewidth]{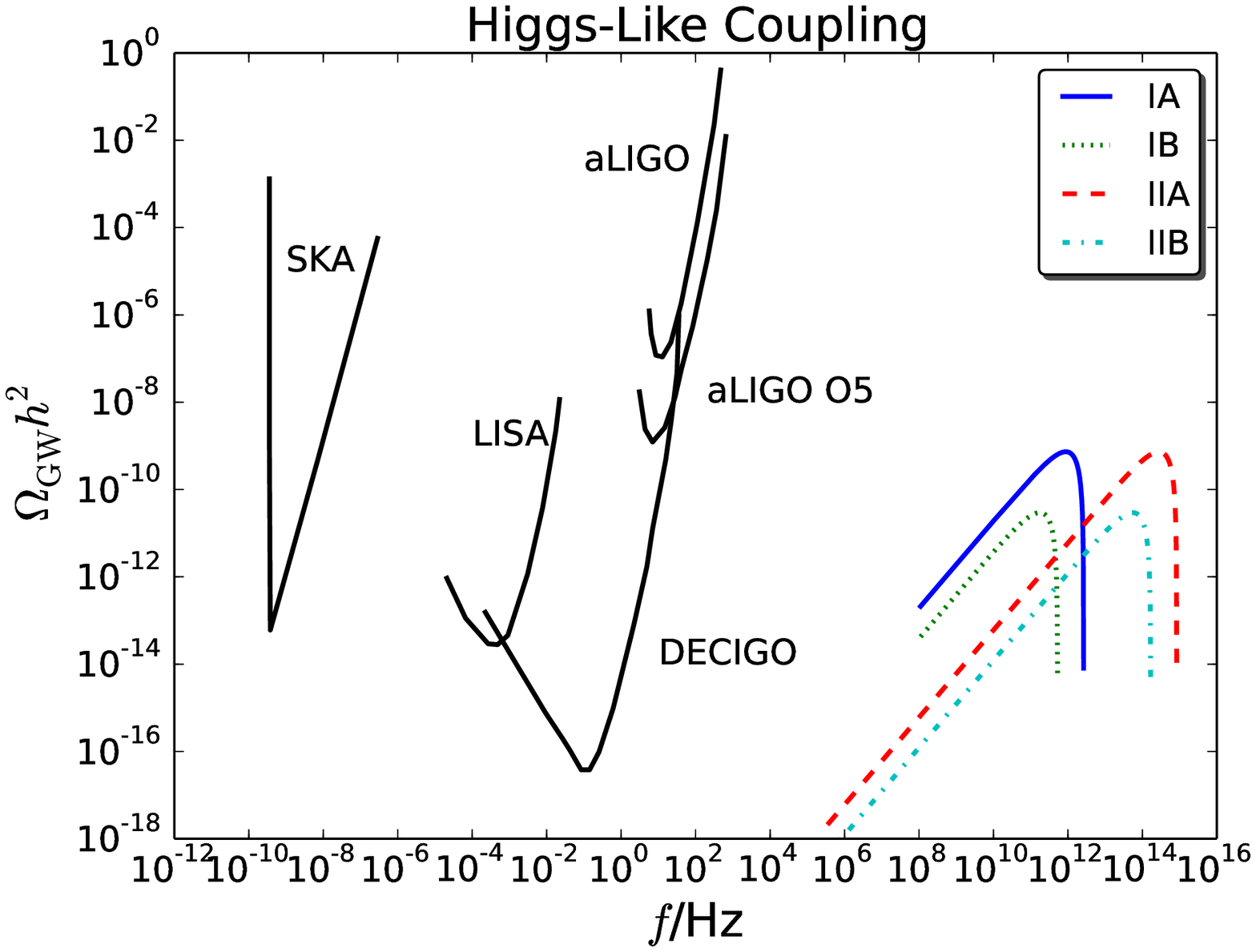}
\includegraphics[width = 0.49 \linewidth]{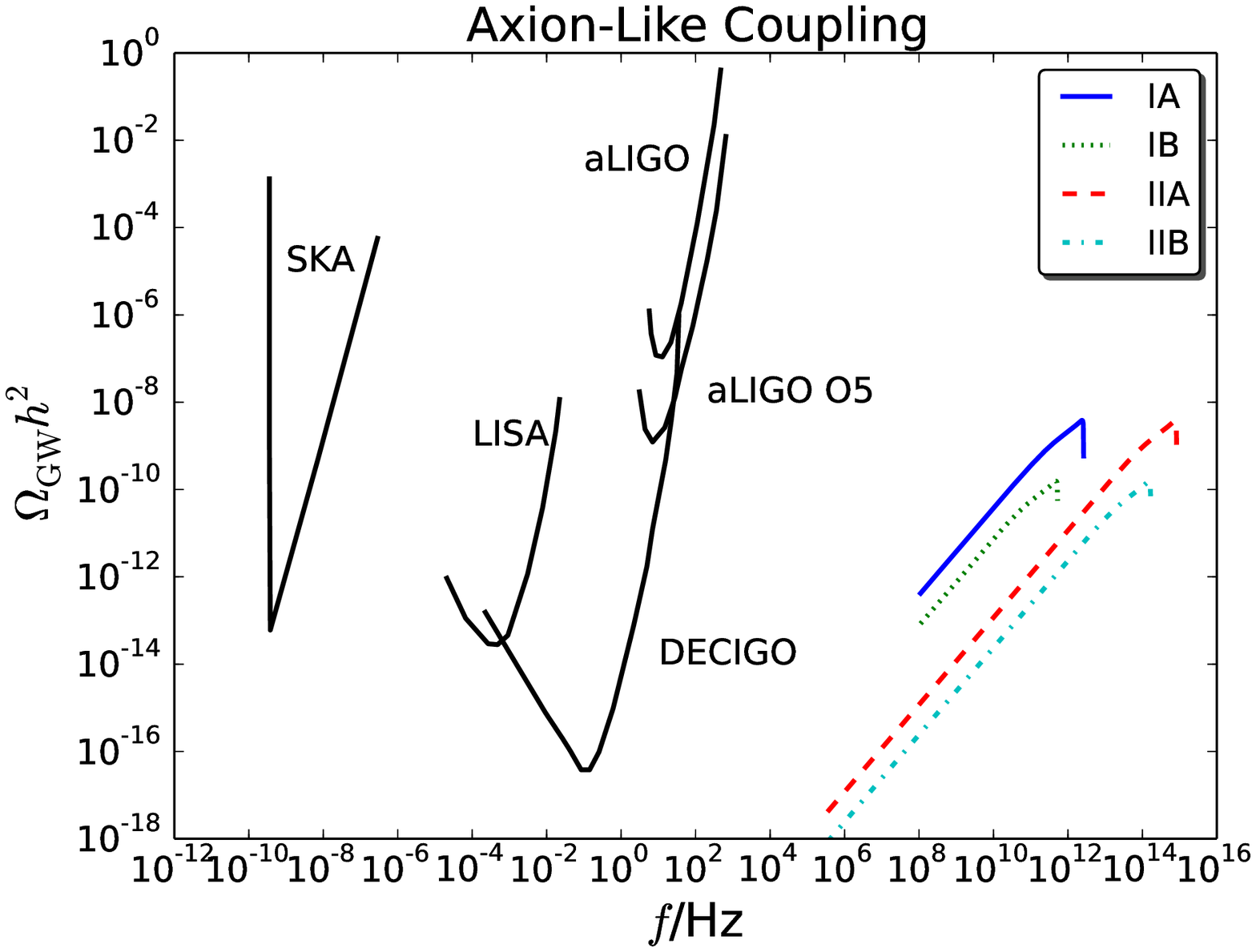}
\caption{The predicted stochastic GW spectra from the inflaton decays for the Higgs-like (left panel) and axion-like (right panel) interactions. Also shown are the sensitivities of several existing and planned GW experiments. See the text for the four choices of model parameters.}\label{FigGWV}
\end{figure}

The predicted GW spectra is displayed in Fig.~\ref{FigGWV} for both the Higgs-like and axion-like interactions, in which we have chosen the following four sets of parameters as in Ref.~\cite{Nakayama:2018ptw}:
\begin{itemize}
\item IA: $M = 0.5 M_{\rm Pl}$, $\Gamma = 10^{-5} M_{\rm Pl}$;
\item IB: $M = 0.1 M_{\rm Pl}$, $\Gamma = 10^{-5} M_{\rm Pl}$;
\item IIA: $M = 0.5 M_{\rm Pl}$, $\Gamma = 10^{-10} M_{\rm Pl}$;
\item IIB: $M = 0.1 M_{\rm Pl}$, $\Gamma = 10^{-10} M_{\rm Pl}$.
\end{itemize}
For the purpose of comparison, we have also presented in both plots sensitivities of the ongoing advanced LIGO~\cite{TheLIGOScientific:2016dpb} and several future GW detection experiments, such as aLIGO designed (aLIGO-O5)~\cite{TheLIGOScientific:2016dpb}, LISA~\cite{Audley:2017drz}, DECIGO~\cite{Kawamura:2011zz} and SKA~\cite{Janssen:2014dka}. It turns out that the two inflaton interactions in Eqs.~(\ref{IntH}) and (\ref{IntA}) lead to qualitatively similar GW spectra for all of the parameter choices, except for some details near the peak frequencies. Moreover, although the amplitudes of predicted GW spectra from the inflaton decays can be large enough, they cannot be probed by GW detectors in the near future since the signal frequencies are too high to be in the detection ranges.

By comparing the four parameter sets, the predicted GW spectra is seen to have interesting dependences on the inflaton mass $M$ and its decay rate $\Gamma$: when $M$ increases with $\Gamma$ fixed, both the amplitudes and peak frequencies of the GW spectra move to the larger values, which is evident by inspecting the cases of IA and IB (or IIA and IIB). In contrast, when $\Gamma$ becomes smaller with $M$ a constant, it is obvious by the comparison of IA and IIA (or IB and IIB) that the GW spectrum only shift to high frequencies while the peak amplitude keeps almost the same. These features can be understood as follows. Apart from several constant factors caused by the cosmic redshift in the GW expression of Eq.~(\ref{GWSpec}), the shape of the GW spectrum is totally determined by the quantity $x^2 d\Gamma_1/[(1-\bar{x})\Gamma dx] $, which has a definite peak at $x_{\rm peak} \sim {\cal O}(0.1)$ for a given inflaton coupling. As is evident in Fig.~\ref{FigVecXbar}, $\bar{x}$ is always much smaller than 1, so that $1-\bar{x} \approx 1$. Thus, as long as $y \ll 1$, $x^2 d\Gamma_1/(\Gamma dx)$ at $x_{\rm peak}$ depends on the inflaton mass as $(M/M_{\rm Pl})^2$ without any reliance on the inflaton decay rate $\Gamma$, which explains the behavior of the GW amplitude as $M$ and $\Gamma$ change. Furthermore, the GW peak frequency can be yielded by $f_{\rm peak}\sim M x_{\rm peak} (T_0/ T_R) \propto M x_{\rm peak} /\sqrt{\Gamma}$, where the first relation is obtained from Eq.~(\ref{xRed}) while the second one from Eq.~(\ref{Tre}). This relation precisely characterizes the peak frequency of the GW spectrum as a function of $M$ and $\Gamma$.

One might wonder what if we modify the inflaton mass $M$ and its decay rate $\Gamma$ so that the shifted GW peak frequency lies around ${\cal O}$(10 Hz) or ${\cal O}$(mHz) which maximizes the sensitivity of advanced LIGO~\cite{TheLIGOScientific:2016dpb} (LISA~\cite{Audley:2017drz}). Unfortunately, no matter how we meet this frequency requirement, the obtained stochastic GW signals are always much smaller than the experimental sensitivity. For example, we can realize the GW peak frequency of ${\cal O}(10~{\rm Hz})$ by tuning the parameters to be $M=1$~GeV and $\Gamma = 0.1$~GeV, the peak amplitude of the GW spectrum for the Higgs-like coupling is as small as $\Omega_{\rm GW}(f_{\rm peak}) \sim {\cal O}(10^{-46})$, which is obviously too tiny to be detected.

As mentioned in Ref.~\cite{Nakayama:2018ptw}, the stochastic GW background might be constrained by the observations of the BBN and CMB since it can contribute to the dark radiation, which is conventionally parametrized by the modification of the effective number of neutrinos $\delta N_{\rm eff}$. For the present GW signal from inflaton decays, by assuming the instantaneous reheating, the contribution to $\delta N_{\rm eff}$ is given by~\cite{Nakayama:2018ptw}
\begin{eqnarray}\label{dNeff}
\delta N_{\rm eff} = \frac{4g_{s}^{R}}{7} \left[\frac{g_s^{\nu\, a}}{g_s^R}\right]^{4/3} \frac{\bar{x}}{1-\bar{x}}\,,
\end{eqnarray}
where $g_s^R$ ($g_s^{\nu\,a} = 43/4$) describes the total degrees of freedom in the SM plasma at the temperature of reheating (just before the neutrino decoupling) as before. If we further assume that the reheating temperature is well above the Electroweak phase transition, all of the SM particles should be relativistic so that $g_s^R = 106.75$. The current measurement of CMB by Planck gives $\delta N_{\rm eff} = 0.085\pm 0.32$~\cite{Ade:2015xua}, while the planned CMB experiments like CMB-S4 can probe its value to the accuracy as $\delta N_{\rm eff} \sim 0.02$--0.03~\cite{Abazajian:2016yjj}. According to Eq.~(\ref{dNeff}), it means that the constraint on $\bar{x} \simeq 10^{-2}$ could be achievable in the future. However, our models predict that $\bar{x}$ cannot be larger than ${\cal O}(10^{-3})$ for both inflaton interactions, even if we push the inflaton mass to the extreme value $M \sim M_{\rm Pl}$ where the present perturbative description is expected to be broken down. Therefore, it seems that for the inflaton decay to vector particles with a gravtion radiated cannot be constrained by dark radiation observations.

\section{Conclusion and Discussion}\label{SecConc}
We have studied the stochastic GW background from the three-body inflaton decays with a pair of vector particles and a graviton in the final states, which is inevitable if the reheating process after inflation is achieved by the perturbative two-body inflaton decays. For two types of inflaton-vector interactions given in Eqs.~(\ref{IntH}) and (\ref{IntA}), we present their respective differential inflaton decay rates of such gravitational decays, from which we notice that the graviton radiation is concentrated at low energies due to the IR divergence. By introducing the IR cutoff scale, we can compute the energy fraction carried by the emitted gravitons, which is found to be as high as $O(10^{-3})$ when the inflaton mass approaches the Planck scale. We have also predicted the stochastic GW background by relating it to the obtained differential inflaton gravitational decay rate, in which we take into account the neutrino decoupling effects ignored previously in Ref.~\cite{Nakayama:2018ptw}. Unfortunately, the obtained GW spectra have been found to be either too high in frequency or too low in amplitude so that they cannot be detected by the ongoing and near-future GW experiments. Thus, the search for such interesting GW signals calls for a new design of high-frequency GW detectors like in Ref.~\cite{Akutsu:2008qv}.

Now we would like to comment on other GW sources in the high-frequency region. One related source is the soft graviton emission~\cite{Weinberg:1965nx} when the plasma particles scatter with each other during reheating. In this case, the typical graviton energy is expected to be in the range $\Lambda < E < T$, in which $\Lambda$ denotes the IR cutoff scale and $T$ the plasma temperature. The differential number density per unit energy $E$ in one Hubble time scale $1/H$ can be estimated as follows~\cite{Weinberg:1965nx,Nakayama:2018ptw}
\begin{eqnarray}
\frac{dn}{dE} \approx \frac{n_\sigma F}{1+F} \frac{1}{E} \frac{{\cal A}}{H}
\end{eqnarray}
where ${\cal A}$ is the reaction rate without graviton emissions, $F$ is the soft graviton emission factor, and $n_\sigma$ denotes the inflaton number density, respectively. Note that if the inflaton is the heaviest particle in the system, it has been argued in Ref.~\cite{Nakayama:2018ptw} that the soft-radiation factor $F$ is dominated by the inflaton scatterings giving $F \approx M^2/(8\pi^2 M_{\rm Pl}^2) $, which is a great suppression for the production of gravitons. Another suppression comes from the factor ${\cal A}/H$, which is anticipated to be smaller than 1. Therefore, soft graviton emissions should be ignorable compared with the inflaton three-body decays.

Another competitive GW source at the high-frequency region is the quantum graviton creation at the end of the inflation~\cite{Ford:1986sy,Peebles:1998qn} as well as during the inflaton coherent oscillations~\cite{Ema:2015dka,Ema:2016hlw}. It was argued in Ref.~\cite{Ema:2015dka,Ema:2016hlw} that such a process could be understood as inflaton annihilations into graviton pairs with a rate $\Gamma_{\sigma\sigma \to hh} \sim H^2 M/M_{\rm Pl}^2$. By comparing with the three-body gravitational inflaton decay rate $\Gamma_1$, we have
\begin{eqnarray}
\frac{\Gamma_1}{\Gamma_{\sigma \sigma \to hh} } \sim \frac{\Gamma_1}{\Gamma} \frac{ \Gamma M_{\rm Pl}^2}{M H^2} \sim \frac{M\Gamma}{H^2}\,,
\end{eqnarray}
where we have used the relation $\Gamma_1/\Gamma\sim M^2/M_{\rm Pl}^2$ which can be seen by our previous discussion. As a result, at the reheating time with $H \sim \Gamma$, this ratio is reduced to $M/\Gamma$ which should be larger than 1 in order for the consistency of the perturbative calculations in the present paper. Therefore, the GWs from the inflaton gravitational decays should dominate over such quantum creations.

Finally, the primordial GW background generated during the inflation could also affect the detection of GWs produced by the inflaton gravitational decays, since the primordial GW spectrum is flat~\cite{Turner:1993vb,Turner:1996ck,Smith:2005mm,Boyle:2005se} up to a high frequency determined by the reheating temperature~\cite{Nakayama:2008ip,Nakayama:2008wy,Kuroyanagi:2008ye}. If there is an overlapping of these two GW spectra, it is possible that the GW from inflaton decays would be buried by the primordial one. As seen in the main text, both the amplitude and the peak frequency of the GW by inflaton decays are proportional to the inflaton mass squared as $M^2/M_{\rm Pl}^2$, so that it is expected that the GWs produced by decays of a heavy inflaton with $M \sim M_{\rm Pl}$ are more optimistic for their detections.

\appendix

\section*{Acknowledgments}
DH would like to thank Dr. Yong Tang for useful discussions on Ref.~\cite{Nakayama:2018ptw}.
DH acknowledges support by the National Science Centre (Poland), research project no. 2017/25/B/ST2/00191.


\end{document}